\documentclass{aa}  
\usepackage{mathtools}
\usepackage{booktabs}
\usepackage[dvipsnames]{xcolor}
\usepackage{BibDef}
\usepackage{natbib}

\usepackage[letterspace=150]{microtype}
\usepackage{kantlipsum}

\usepackage{graphicx}
\usepackage{txfonts}

\begin{document}

   \title{{Revisited mass-radius relations for exoplanets below $120 M_{\oplus}$}}

   \author{J.F. Otegi\inst{1,2} 
          \and
          F. Bouchy\inst{2}
          \and
          R. Helled\inst{1}
          }

   \institute{Institute for Computational Science, University of Zurich,
              Winterthurerstr. 190, CH-8057 Zurich, Switzerland\\
              \email{jonfr17@gmail.com}
         \and
              Observatoire Astronomique de l’Universit\'e de Gen\`eve, 51 Ch. des Maillettes, – Sauverny – 1290 Versoix, Switzerland\\
             }

 
  \abstract
  { The masses and radii of exoplanets are fundamental quantities needed for their  characterisation. Studying the different populations of exoplanets is important for  understanding the demographics of the different planetary types, which can then be linked to  planetary formation and evolution. We present an updated exoplanet catalogueue based on reliable, robust, and, as much as possible accurate mass and radius measurements of transiting planets up to 120 $M_{\oplus}$. The resulting mass-radius (M-R) diagram shows two distinct populations, corresponding to rocky and volatile-rich exoplanets  which overlap in both mass and radius. The rocky exoplanet population shows a  relatively small density variability and ends at mass of  $\sim25 M_{\oplus}$, possibly indicating the maximum core mass that can be formed. We use the composition line of pure water to separate the two populations, and infer two new empirical M-R relations based on this data: $ M = (0.9 \pm 0.06) \ R^{(3.45 \pm 0.12)}$ for the rocky population, and  $ M = (1.74 \pm 0.38) \ R^{(1.58 \pm 0.10)}$ for the volatile-rich population.  While our results for the two regimes are in agreement with previous studies, the new M-R relations better match the population in the transition region from rocky to volatile-rich exoplanets, which correspond to a mass range of 5-25 $M_{\oplus}$, and a radius range of 2-3 $R_{\oplus}$. \\

  \vspace{15mm}
  }
   \maketitle
%

\section{Introduction}

To date, more than 4000 exoplanets have been discovered. The Kepler mission has clearly impacted the field with the detection of more than 2300 exoplanets. For many of the Kepler exoplanets, radial velocity follow-up is restricted to a small fraction corresponding to the brightest host stars. As a result, in order to characterise the exoplanets researchers often rely on a theoretical mass-radius (hereafter M-R) relation. Knowledge of both the planetary mass and radius allows us to estimate the planetary bulk density and infer the possible compositions and internal structures.  In addition, the M-R relation is used to explore the demographic of exoplanets in a statistical sense. These demographics can then be linked to the physical and chemical processes driving planet formation and evolution, such as the planetary mass function, primordial atmosphere mass, migration, atmospheric loss, inflation mechanism, etc. providing constraints on the formation models. \\

Various studies have been dedicated to the investigation of the internal structures of exoplanets  \cite[e.g.][]{2007Icar..191..337S,2007ApJ...669.1279S,2014ApJhowe,dorn-2017,Lozovsky_2018} and to the investigation of 
the M-R relation of exoplanets of different populations.  
Parametric models,  power laws in particular, have been proposed to fit the M-R relation. These are typically empirical relations based on exoplanet data found in the main exoplanet catalogues:  Extrasolar
Planets Encyclopaedia\footnote{exoplanet.eu} \cite[][]{weiss2013,weiss_marcy2014,bashi2017} or NASA Exoplanet Data Archive\footnote{exoplanetarchive.ipac.caltech.edu} \citep{woflgang2015}.
As shown in \cite{bashi2018}, despite the overall good agreement between these catalogues, there are also some differences. On the other hand,  the Extrasolar
Planets Encyclopaedia has the largest coverage of exoplanets, probably due to its less restrictive  selection criteria. 
On the other, the NASA Exoplanet Database  has a 'removed targets' list, providing a more rigorous selection process, and is the most updated catalogue. Recently, \cite{zeng2016} inferred a   semi-empirical M-R relation depending on the core mass fraction, followed by a detailed forecasting model using a probabilistic M-R relation using Markov Chain Monte Carlo (MCMC).  \citep{chen-kipping2017}. 

While such studies are crucial for a more detailed characterisation of exoplanets, it should be noted that the number of discovered exoplanets increases rapidly and the estimates for the masses and radii are continuously being updated. In addition, one has to account for the fact that some of the listed mass and/or radius determinations are not reliable, which can affect the inferred conclusions (see Section 2 for details).  
In this study, we go through the entire NASA Exoplanet catalogue, in order to create a 'filtered' sample of exoplanets with robust and reliable mass and radius measurements. We consider exoplanets with masses below $120 M_{\oplus}$ in order to focus on the transition between small-size terrestrial planets and the population of giant gaseous planets. 
We use our updated catalogue to describe the properties of two distinct populations: rocky and volatile-rich population. We derive updated M-R relations for these two populations and investigate the dependence of the M-R diagram with other external parameters.

\section{{Exoplanet selection with reliable measurements of mass and radius}}

 We use the NASA Exoplanet Archive from June 2019 as a starting point  since it is the most up-to-date catalogue, and, in addition, it provides access to all the references relevant for a given observed exoplanet. 
  We build a "reliable and updated" catalogue by applying the following selection criteria:
 
 \begin{figure*}[h]
\centering
  \begin{tabular}{@{}cc@{}}
    \includegraphics[scale=0.42]{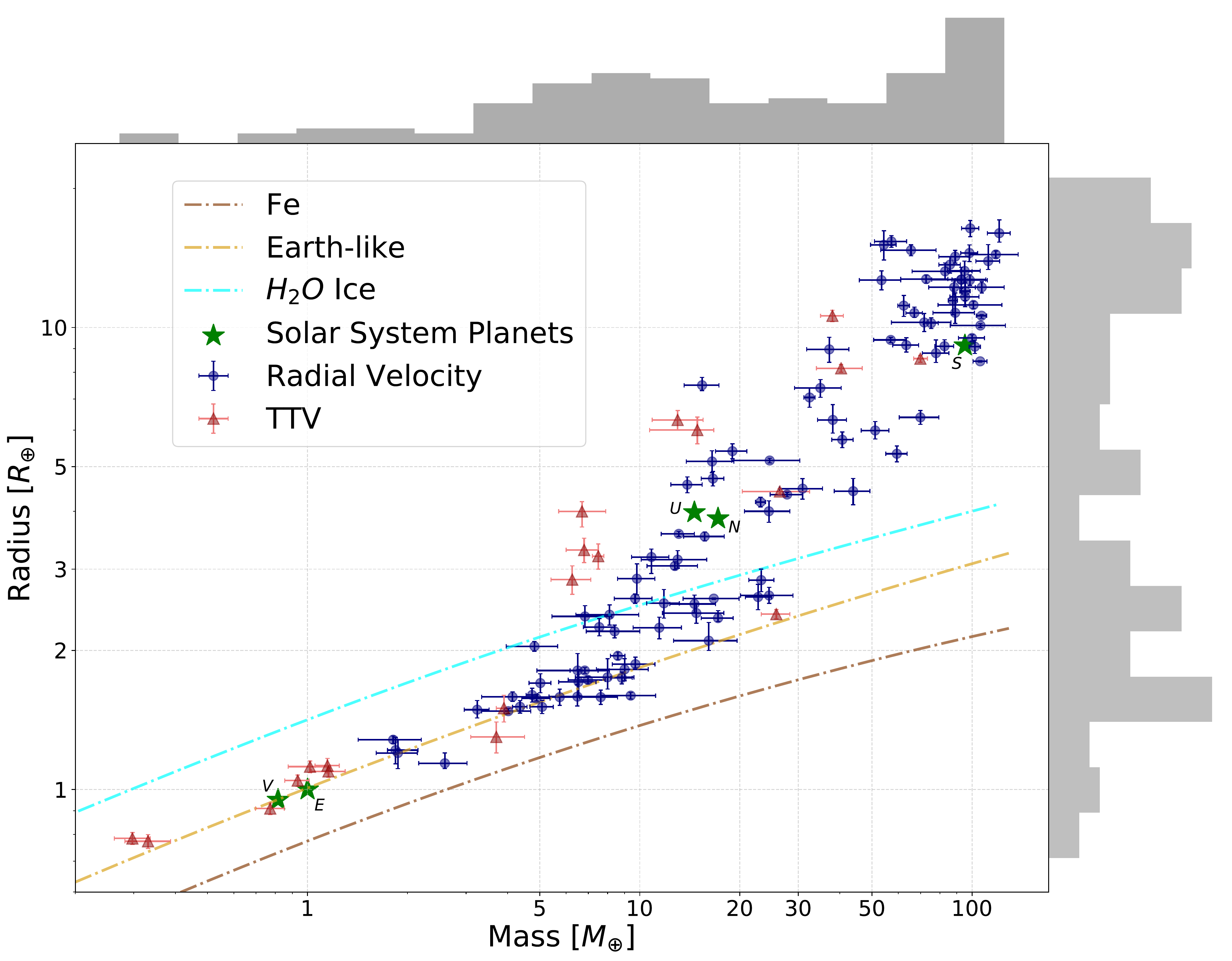}
  \end{tabular}
  \caption{ Revisited M-R diagram after applying our criteria to keep  reliable and robust mass measurements with relative uncertainties smaller than 25\% for mass and smaller than 8\% for radius. The red triangles and blue circles correspond to data with mass determination from TTVs and RVs, respectively. We also  display the composition lines of pure-iron (brown), Earth-like planets (light-brown) and water ice (blue) \cite[][]{Dorn2015}. We also plot the contour lines and the distribution of exoplanet mass (top) and radius (right) of our sample.}  
\end{figure*}

 \begin{figure*}[h]
\centering
  \begin{tabular}{@{}cc@{}}
    \includegraphics[scale=0.5]{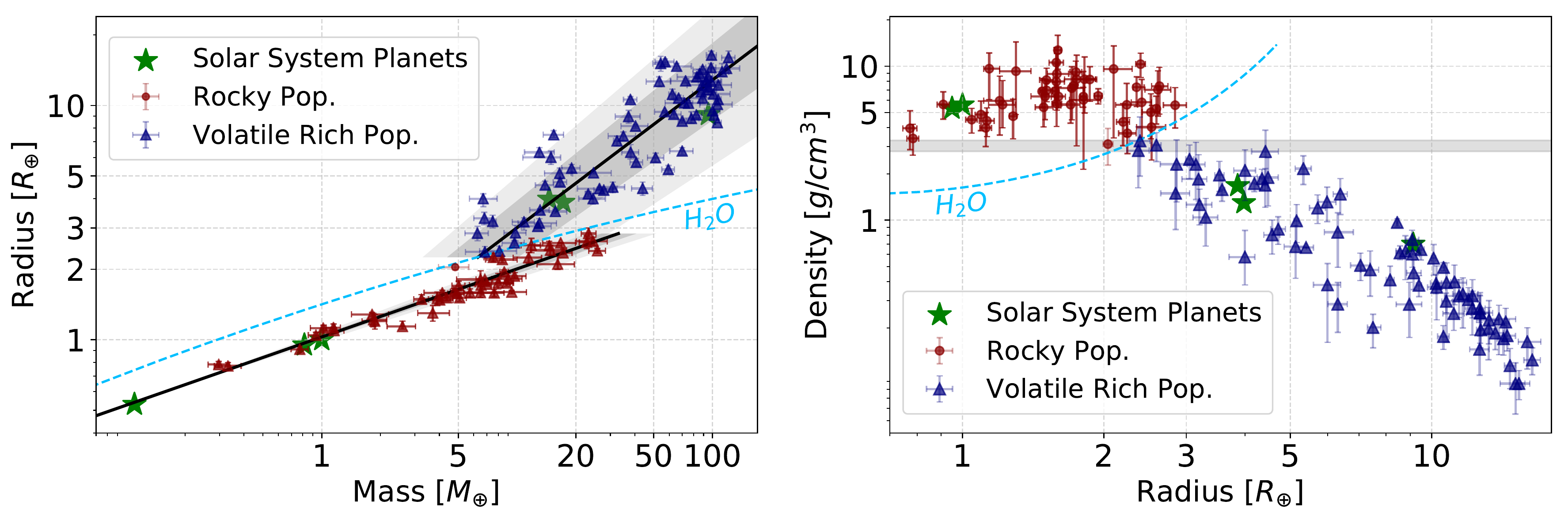}
  \end{tabular}
  \caption{Left: M-R relations fitting rocky and  volatile-rich  populations. Dotted line corresponds the composition line of pure water using QEOS for a temperature of 300K \cite[][]{More1988}. The grey and light-grey envelopes represent the $\pm 1 \sigma $ and $\pm 2 \sigma $ regions of the fit.  Right: Density against radius for our catalogue. Rocky and the volatile-rich populations are separated by the composition line of pure water \cite[][]{Dorn2015}. The grey envelope indicates the region between $2.8~g \, cm^{-3}$ and $3.3~g \, cm^{-3}$. } 
\end{figure*}

 \begin{figure*}[h]
\centering
\begin{tabular}{@{}cc@{}}
    \includegraphics[scale=0.5]{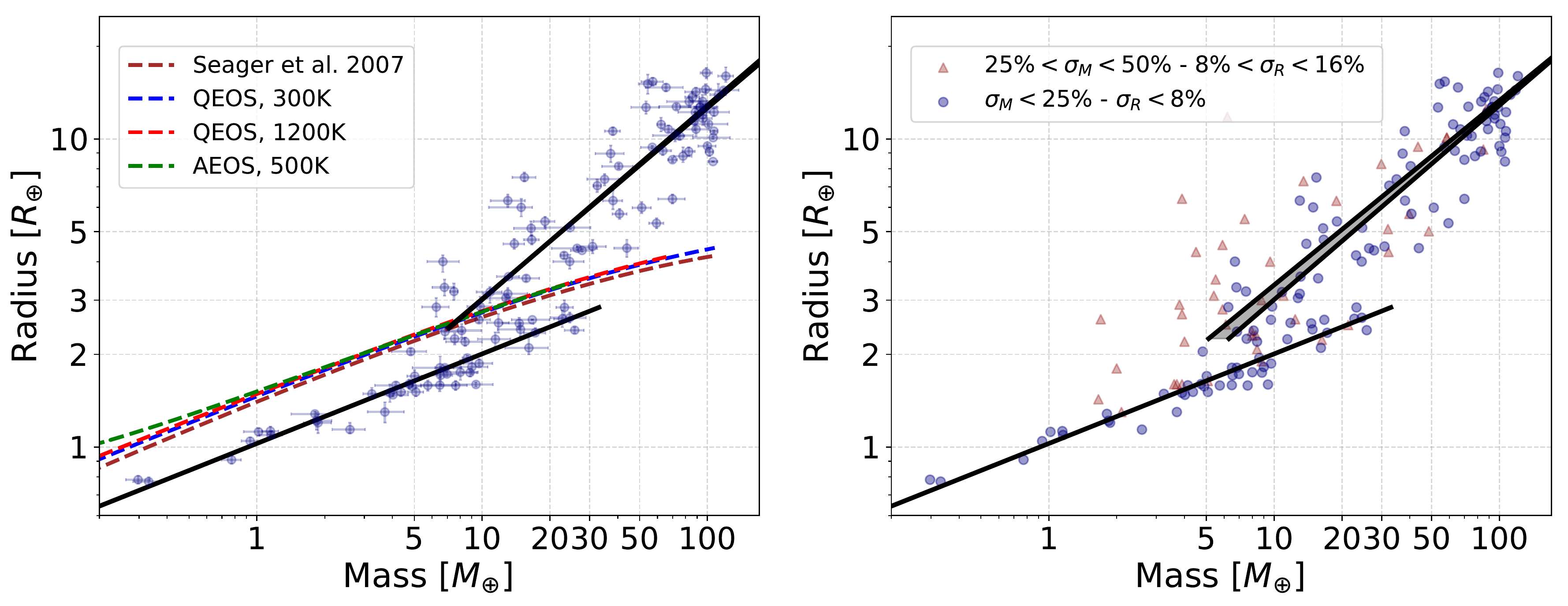}
  \end{tabular}
  \caption{Right: M-R diagram comparing obtained M-R relations when using different EOS for water: the polytropic EOS of \cite{Seager2007} (brown), QEOS assuming temperatures of 300K and 1200K \cite[][]{More1988}, and ANEOS \cite[][]{Thompson1990} with a surface temperature of 500K.
   Right: M-R diagram comparing obtained M-R relations when using different cuts for the mass and radius uncertainties when building the revisited catalogue:  $\sigma_M / M = 25 \%$ and $\sigma_R / R=8 \%$, and $\sigma_M / M= 50 \%$ and $\sigma_R / R =16 \%$. The grey envelope corresponds to the difference between the two. Blue circles represent the exoplanets with  $\sigma_M / M= 25 \%$ and $\sigma_R / R =8 \%$, and red triangles represent the planets with uncertainties between  $\sigma_M / M= 25 \%$ and $\sigma_R / R =8 \%$, and $\sigma_M / M = 50 \%$ and $\sigma_R / R=16 \%$.}
\end{figure*}

\begin{enumerate}[a)] 
\item We selected the data from the NASA Exoplanet Archive  from July 2019 for planets with masses up to $120M_{\oplus}$ 
and filtered the data to consider only exoplanets with measurement uncertainties smaller than $\sigma_M/M=25 \%$ $\sigma_R/R=8 \%$. These thresholds correspond to the median uncertainty and make it possible to have the same impact on the density uncertainty. 
\item We added the mass measurements of the exoplanets orbiting around Trappist-1 from \cite{grimm2018},  who used new K2 transit light curves to recompute the masses through TTVs and shrank the mass uncertainties from $30\%-95\%$ to  $5\%-12\%$. 
\item We discarded the mass determinations inferred by \cite{stassun2017}, where the host star masses and radii were replaced by the value derived from GAIA photometry and with uncertainties clearly  overestimated. These revised values affect the planetary mass estimation. Therefore, in the cases in where the NASA Exoplanet Archive selects this study as the reference paper, we replace them with the most updated mass estimate (Kepler-78b,Kepler-93b, CoRoT-7b, Kepler-454b, HD 97658b, HIP 116454b, WASP-29b, WASP-69b, WASP-117b, HD 149026b, WASP-63b) \footnote{References used for these planets are listed in Table A.1}. 

\item In some cases \cite{Marcy2014} give an estimate of the planets masses for single transiting planet with a weak level of validation/confirmation.  Several mass estimates are based on very few radial velocity data points with underestimated uncertainties. We therefore did not use mass estimates given by \cite{Marcy2014} for exoplanets with non-robust measurements (Kepler-406c, -97b, -98b, -102b, -48b, -99b, -406b, -100b, -48b, -96b, -102e, -25b, 103b, -106c, -106e, -113b, -103c). These exoplanets do not have any other mass estimates  from other studies with measurement uncertainty smaller than $\sigma_M/M=25 \%$ $\sigma_R/R=8 \%$, and therefore were not included in our catalogue. 

\item The TTV measurements reported by \cite{xie2014} differ significantly from the mass measurements reported by other groups \cite[][]{Hadden2014,HaddenLithwick2017}.  In addition, several of their mass and radius estimates imply that several exoplanets with masses greater than $30 M_{\oplus}$ are denser than pure-iron (e.g., Kepler-128b and Kepler-128c, for which \cite{HaddenLithwick2017} estimated masses below $1 M_{\oplus}$). Therefore we also  excluded the mass estimates provided by this study. 
\item \cite{HaddenLithwick2017} provide the planetary masses through TTVs for 150 Kepler exoplanets, which are not used in the NASA Exoplanet Archive. In addition, they introduce a robustness criterion for TTVs, and consider that only 50 out of 150 mass measurements are reliable. We relied on their robustness criterion and discarded the unreliable TTVs mass determination. 
\item We updated some mass measurements to the ones presented in more recent publications (Kepler-10b, Kepler-65d,  GJ 9827b, 55-Cnc e, K2-55b K2-261b, HAT-P-18b, HAT-P-12b, WASP-20b)$^3$ and we included several exoplanets that are missing from the NASA Exoplanet Archive (Gl-357b, HD 39091c, HD 3167b, K2-131b, HD 15337c, HD 213885b, EPIC 220674823b, HD 3167c, K2-180b, K2-24c, GJ 143b, HD 21749b, WASP-166b, WASP-107b, HAT-P-48b, HAT-P-47b, Kepler-425b, NGTS-5b, HATS-43b, WASP-160b, Kepler-427b, WASP-181b, K2-295b, EPIC 220501947b, Kepler-426b, Qatar-8b)$^3$. 
\end{enumerate}

Figure 1 shows the M-R diagram after applying the  aforementioned selection criteria. It also shows the mass and radius histograms of the exoplanets in our sample.  Intermediate steps of the data selection are shown in the appendix in Figure A.1. The planets that are included  in our "filtered" catalogue are listed in the appendix  (Table A.1), where we also provide the references used by the NASA Exoplanet archive and the ones used in this work. 
It should be noted that as other catalogues, also ours suffers from observational biases and is  incomplete. As a result, it cannot be used to make conclusions about the planetary occurrence rates. 
Our revised catalogue of transiting planets below $120M_{\oplus}$ is accessible on the Data \& Analysis Center for Exoplanet  DACE\footnote{dace.unige.ch}.  

 \begin{figure*}[h]
\centering
  \begin{tabular}{@{}cc@{}}
    \includegraphics[scale=0.5]{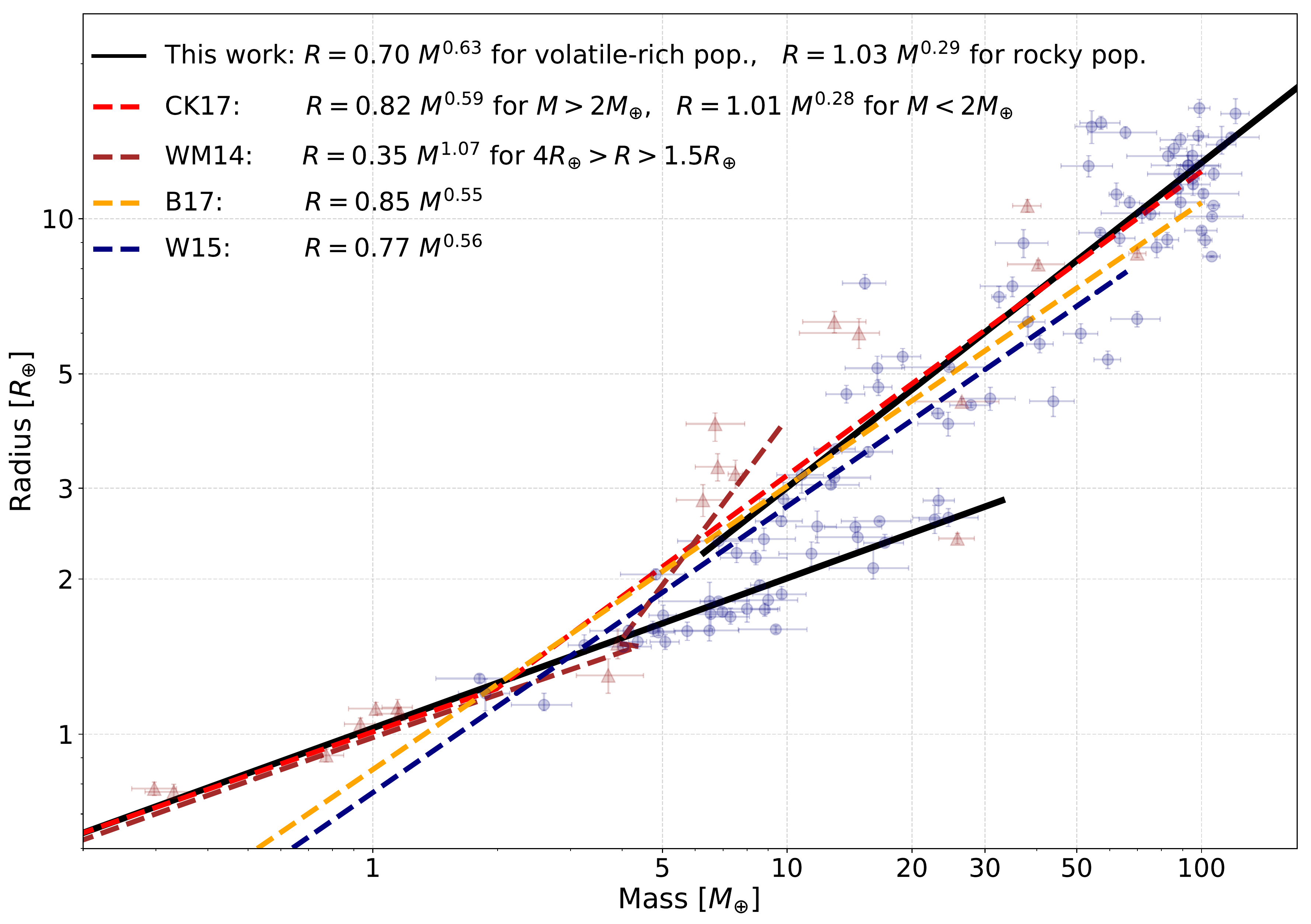}
  \end{tabular}
  \caption{Comparison of M-R relations in the literature with the one obtained from our revisited catalogue. 
  Red triangles and blue circles correspond to our revised catalogue with mass determination from TTVs and RVs, respectively. The analytic expressions of the M-R relations CK17, WM14, B17 and W16 corresponds to \cite{chen-kipping2017}, \cite{weiss_marcy2014},  \cite{bashi2017} and \cite{woflgang2015}, respectively.}
\end{figure*}

\section{Analysis of the revisited M-R diagram}
\subsection{Two distinct exoplanet populations}

Our revisited M-R diagram clearly shows two distinct exoplanet populations: one of them closely follows an Earth-like composition, and a second one corresponds to a more volatile-rich composition. 
It is important to note that even when an exoplanet lies in the M-R  diagram of an Earth-like composition, its actual  relative composition of iron, silicates and water  could be different, given the degeneracy of the problem.

Nevertheless, the amount of water or H/He envelope of  these exoplanets is expected to be small in comparison to the refractory materials (e.g. silicates, iron). Therefore, it is reasonable to assume that these exoplanets are mostly rocky\footnote{in this study 'rocky exoplanets' refer to exoplanets that  mostly consist of  metals and rocks}. This population follows the Earth-like composition up to a mass of $\sim 25 M_{\oplus}$ including Kepler-411b. However,  between 10 and 25 $M_{\oplus}$, exoplanets appear to be slightly less dense than the ones following the Earth-like composition. These objects might be ice-rich cores, but are unlikely to be volatile-rich.  
Therefore we include them in the rocky population, which contains 'naked-cores' up 10 $M_{\oplus}$ and slightly more ice-rich exoplanets from 10 to 25 $M_{\oplus}$.  This suggests that this upper limit corresponds to the maximum core mass that can be formed, and it is important to note that this region of the diagram does not suffer from observational biases since heavier planets are easy to detect in radial velocity. This estimate of the maximum core mass is in fact consistent with theoretical calculations of giant planet formation, and with the estimated core masses of the giant planets in the solar system: for Jupiter structure models typically infer core masses between $7 M_{\oplus}$ and $25 M_{\oplus}$ \citep[e.g.][]{Guillot2017,Wahl2017,Helled2017}, and Saturn's core mass is expected to be of the order of $20 M_{\oplus}$ \citep[e.g.][]{Saumon2004,Iess2019}. 

In addition, giant planet formation models with pebble accretion estimate the pebble isolation mass to be between $10 M_{\oplus}$ and $20 M_{\oplus}$ \citep[e.g.][]{Johansen2017,Bitsch2019}, which is also consistent with our estimated maximum core mass of the order of $25 M_{\oplus}$. This result suggests that our M-R relation can be used to confirm and test theoretical predictions. 

It is interesting to note that none of the exoplanets in our sample is found to be consistent with a pure iron composition. 
A structure model of highest density planet Kepler-107c ($\bar{\rho} = 12.65~g \, cm^{-3}$) suggests that it has  a large iron core and a silicate mantle, corresponding to 70\% and 30 \% of the planetary mass, respectively  \citep[e.g.][]{Bonomo2019}.     \\


 The second population shown in our revisited M-R diagram corresponds to less dense planets with a more volatile-rich composition. The density-radius diagram displayed on the right panel of Figure 2 makes it possible to distinguish the two populations separated by the composition line of pure water (see section 3.2 for details).  The rocky population presents a nearly flat density up to 2-3 $R_{\oplus}$,  corresponding to  behaviour of exoplanets made of refractory materials. The volatile-rich population shows a decreasing density from 2-3 $R_{\oplus}$ to 12 $R_{\oplus}$. 


Although the M-R is biased toward lower masses, it seems that $5 M_{\oplus}$, like HD39091, is the lower limit from which an exoplanet can accrete and maintain a gaseous envelope. However, it is possible that the data sample is incomplete and suffers from observational biases, since for a given radius it is much easier to detect more massive planets. 

The dispersion of the volatile-rich population is  significantly larger than the one in the rocky population.  It may reflect different core masses accreting gas. Another reason could be that volatile rich exoplanets are very sensitive to insolation, and therefore the dispersion reflects exoplanets with different stellar irradiations (this hypothesis is further discussed in Section 3.2).\\

\cite{Fulton2017} find that at small radial distances there is a lack of planets with radii between $1.5 R_{\oplus}$ and $2 R_{\oplus}$, known as the Fulton Gap, suggesting a transition between the super-Earth and sub-Neptune populations.
Nevertheless, as discussed above, our revisited M-R diagram shows two exoplanet populations with a large overlap in both mass and radius. The overlap in mass ranges between 5 and 25 $M_{\oplus}$, and in radius from 2 to 3 $R_{\oplus}$. 
Although planets smaller than 1.8 $R_{\oplus}$ are clearly part of the rocky population, planets with a larger radius could belong to both populations. Therefore the planetary mass or radius alone cannot be used to distinguish between the two populations. We therefore use the composition line of pure water to separate the rocky and volatile-rich populations. This provides a more physical criterion to divide the populations as planets that sit above the pure-water line are expected to consist of volatile materials (e.g. H-He).


 \begin{figure*}[h]
\centering
  \begin{tabular}{@{}cc@{}}
    \includegraphics[scale=0.45]{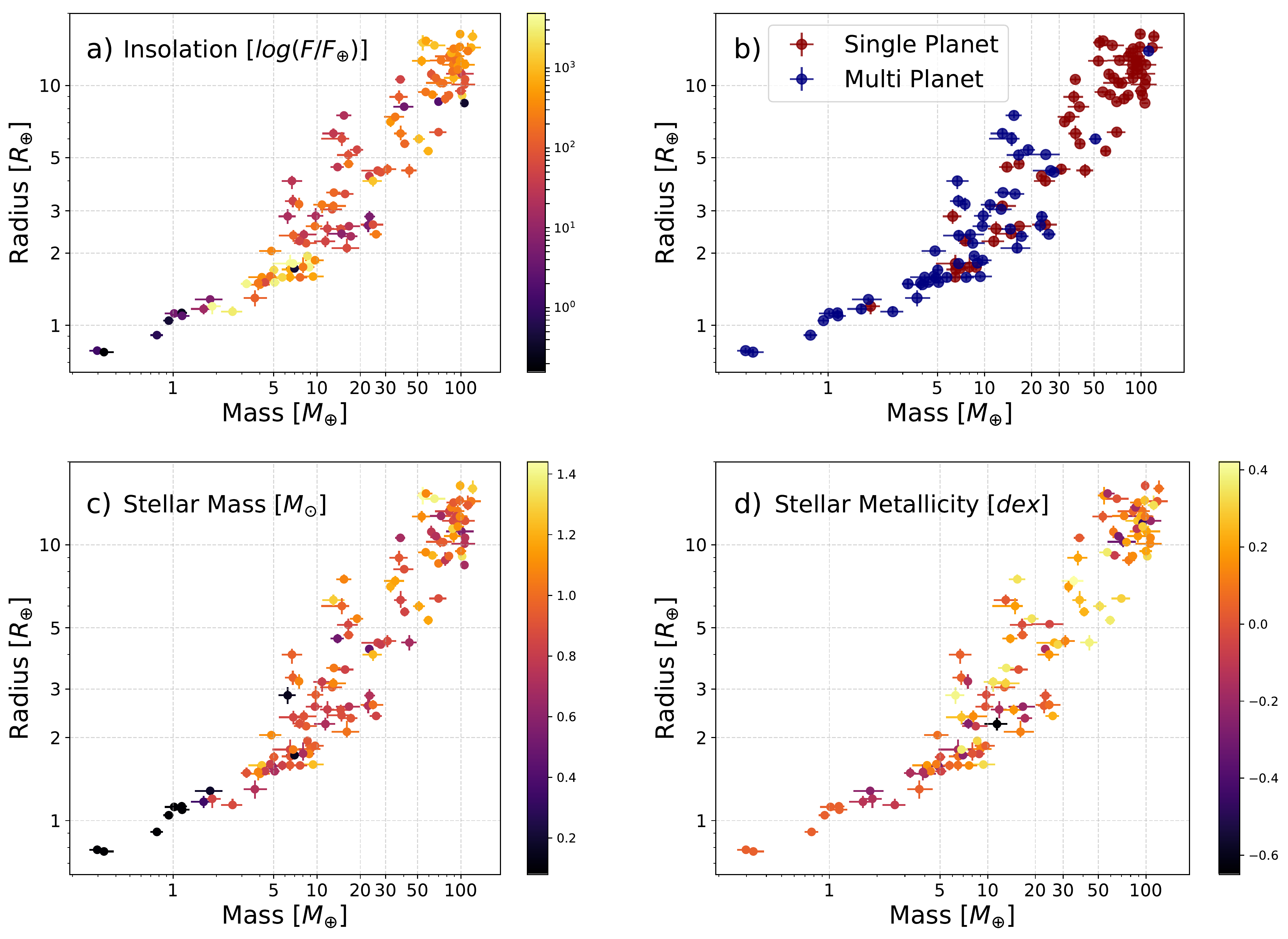}
  \end{tabular}
  \caption{M-R diagrams from our revisited catalogue showing dependence with insolation (a), multiplicity (b), stellar mass (c) and stellar metallicity (d).}
\end{figure*}

\subsection{The Mass-Radius relations}

Published M-R relations to date have divided the different exoplanet populations using mass cutoff \cite[e.g.][]{chen-kipping2017} or radius cutoff \cite[e.g.][]{weiss_marcy2014}. In this study, we divide the super-Earth and volatile-rich populations using the composition line of pure-water. We fit the M-R relation of the two populations using a total least squares method, in which observational errors on both dependent and independent variables are considered. Figure 2 (left) shows the inferred M-R relations for the two exoplanet populations,
assuming an M-R dependence of $R=AM^B$. The results of the fit are shown in Equations 1 and 2: 
\begin{equation}
  \ \ \ \ \ \ \ \ \ \ \ \ \ \ \ \  R=
    \begin{cases}
      (1.03 \pm 0.02) \ M^{(0.29 \pm 0.01)} &, \ if \ \rho > 3.3~g \, cm^{-3} \\
      (0.70 \pm 0.11) \ M^{(0.63 \pm 0.04)}& , \ if \ \rho < 3.3~g \, cm^{-3} \\
    \end{cases}       
\end{equation}
or 
\begin{equation}
  \ \ \ \ \ \ \ \ \ \ \ \ \ \ \ \  M=
    \begin{cases}
      (0.90 \pm 0.06) \ R^{(3.45 \pm 0.12)} &, \ if \ \rho > 3.3~g \, cm^{-3} \\
      (1.74 \pm 0.38) \ R^{(1.58 \pm 0.10)}& , \ if \ \rho < 3.3~g \, cm^{-3} \\
    \end{cases}       
\end{equation}


The rocky exoplanet population presents a relatively small dispersion around the M-R relation reflected by the small uncertainties on the fitted parameters. On the other hand, the volatile-rich population  presents a larger dispersion around the adjusted relation reflecting a larger diversity in composition. Contrary to  previous studies, all the   observed masses and radii are found to be at less than $2 \sigma$ of our M-R relations. In addition, the Solar System planets with masses smaller than $120 M_{\oplus}$ lie on the derived M-R relations (except Mercury, which is anomalously dense).\\

Figure 2 (right) shows the density-radius diagram of our planetary catalogue, and a density-mass diagram is included in the appendix. 
Another physically-motivated approach is to divide the rocky and volatile-rich populations by using a density-cutoff. This density cutoff should be between the minimum density of a rocky planet and the maximum density of a volatile-rich planet. The lowest density exoplanets that are expected to be rocky in our sample are the Trappist planets, with $M_{solid}/M_{total} > 84\%$ and $R_{gas}/R_{total}< 2 \%$ \cite[e.g.][]{Dorn2018}. This sets a maximum value for the density-cutoff of 3.3 $g \, cm^{-3}$. On the other side, K2-55b is the densest planet in the volatile-rich population and has an estimated envelope mass fraction of 12$\%$ \cite[e.g.][]{Dressing2018}. It then sets a minimum limit for the density-cutoff of 2.8 $g \, cm^{-3}$. The grey envelope in Figure 2 (right) shows the region between $2.8~g \, cm^{-3}$ and $3.3~g \, cm^{-3}$. Using a physically-motivated density cutoff to divide the two populations, or dividing the two populations by the composition line of pure water, lead to nearly identical results. This suggests that the physically-motivated approach of dividing the two populations using the pure-water curve is essentially equivalent to the division of the populations by a density cutoff of $\sim 3~g \, cm^{-3}$.  Nevertheless, the pure-water composition curve is less arbitrary and is based on physical arguments. \\

 The composition line of pure-water depends on the EOS (Equation of State) used and the planetary temperature. Figure 3 (left) shows how the derived M-R relations depend on these choices, comparing the polytropic EOS of \cite{Seager2007}, QEOS, assuming surfaces temperatures of 300K and 1200K \cite[][]{More1988}, and ANEOS, for a  temperature of 500K. We find that the results are insensitive to the water EOS used and the assumed temperature.  It is also important to note that the pure-water composition line also depends on the assumed pressure, which here was assumed to be one bar. This assumed pressure corresponds to a water world without a water vapor atmosphere, and, therefore, the M-R relation represents a lower bound for the radii of pure water planets. Nevertheless, we find that the coefficients of the M-R relations are relatively insensitive to the assumed surface pressure. \\

 In order to investigate the impact of mass and radius uncertainties to our M-R relations, we extended our catalogue to planets with uncertainties two times larger in both mass (50\%) and radius (16\%). 
 The right panel of Figure 3 shows a comparison of the M-R relations obtained  using cuts of $25 \%$ and $50 \%$ for the mass uncertainty, and $8 \%$ and $16 \%$ for the radius uncertainty. The change of uncertainty has no  significant impact on the M-R relations for both the rocky and volatile-rich populations and the adjusted parameters are within error bars. 
 We can therefore conclude that our fit does not depend on the limit in mass and radius uncertainties  and is not significantly biased by our selection criteria.
 For the rocky population, the M-R relation indicates that the bulk density is nearly constant. For volatile-rich populations, the bulk density scale with M$^{-1}$. \\

 Figure 4 compares the M-R relations in the literature with the one obtained from our revisited catalogue. Our M-R relation is similar to the one inferred by \cite{chen-kipping2017}, but the transition from the rocky to the volatile-rich regime  is defined for a mass  of $2 M_{\oplus}$, so they underestimate the masses of most of the rocky exoplanet population. Our inferred  transition mass between rocky and volatile-rich planets is found to cover a large range of masses  ($10-25M_{\oplus}$) and is significantly higher than the $2 M_{\oplus}$ predicted by \cite{chen-kipping2017}. Our relation is also very  similar to the one  derived by \cite{weiss_marcy2014} for planets smaller than $R=1.5 R_{\oplus}$. For larger radii their fit differs from all the relations in the literature. The relations in \cite{bashi2017} and \cite{woflgang2015} are relatively close to our relation for the volatile-rich population, but they use  a single and unique relation for all the planets and do not represent the rocky population correctly.  \\

\subsection{Dependence on other parameters}

The exoplanets in our sample with masses measured via TTV are statistically less massive than the ones measured by RV. As discussed in \cite{Steffen2016}, the sensitivity of TTVs and RVs can be expressed by: 

\begin{equation}
     SNR_{TTV} \sim \frac{M_p R_p^{3/2} P^{5/6}}{\sigma _{TTV} } \ \ \ \ \ \ \ \ \ , \ \ \ \ \ \ \ \ \ SNR_{RV} \sim \frac{M_p}{\sigma _{RV} P^{1/3}} \ \ ,
\end{equation}
where $\sigma$ is the intrinsic uncertainty of a measurement. This is a clear observational bias since it is easier for the RV technique to derive the mass of a short period planet, while the TTV technique can determine masses more easily for longer periods.  Exoplanets orbiting close to their host stars may have smaller atmospheres (lost through evaporation) and therefore, higher densities. In fact, we can see that the discrepancy  between TTVs and RVs arises in the volatile-rich population, while in the rocky population, masses measured by TTVs and RVs overlap. Therefore, for  volatile-rich exoplanets with densities smaller than 3.3 $g \, cm^{-3}$, the RV method tends to detect more massive exoplanets, while the sensitivity of the TTVs seems to be more uniform. 
The fact that our exoplanet sample is dominated by RV measurements and that the current TTVs estimates are significantly less accurate and less robust is likely to bias the demography of exoplanets.

It is interesting to investigate whether there is any trend with a third parameter. Figure 5 shows the dependence of M-R diagram with insolation (a),  multiplicity (b), stellar mass (c), and stellar metallicity (d).  
The densest exoplanets of the rocky population (removing the Trappist-1 exoplanets and the ones beyond $M=10  M_{\oplus}$) are noticeably more irradiated than the population of exoplanets with a volatile envelope  although there is no clear difference between these two populations in terms of stellar mass. It suggests that strongly-irradiated exoplanets (with  insolation  greater than $1000 F_{\oplus}$) are rocky, probably because their H/He envelopes have been photoevaporated by the high-energy radiation from the host star  \cite[e.g.][]{owenlu2013,lopez-fortney2013,jin2014,zeng2017,jin-mordasini2018}.\\

When comparing the population of single exoplanets and multi-planetary systems, we see that the majority of exoplanets in the low-mass regime ($< 25 M_{\oplus}$) are multi-planets, while beyond $30 M_{\oplus}$, almost all of them are single, although this could be a result of an observational bias. We do not find clear trends with stellar mass and metallicity. We suggest that more data and a  systematic analysis of these results, including the  observational biases, are required in order to understand the relation between planetary and stellar properties.


\section{Conclusion}

We present an updated exoplanet catalogue based on reliable and  robust  mass and radius measurements up to $120 M_{\oplus}$, which is available in the DACE  platform\footnote{dace.unige.ch}. The resulting M-R diagrams clearly shows two distinct populations, corresponding to rocky exoplanets and volatile-rich exoplanets. The rocky exoplanet population shows a  relatively small density variability and ends at a mass of  $\sim25 M_{\oplus}$, possibly indicating the maximum core mass that can be formed.  \\

We present new empirical M-R  relations based on this catalogue. Since the two exoplanet populations overlap in mass and radius, we divide the rocky and  volatile-rich regimes by the composition line of pure water and fit both populations. We show that the coefficients we get are rather insensitive to the used composition line of pure water and the limits on  mass and radius  uncertainties chosen for the catalogue.  We compare our M-R relations with previous published ones and we identify their limitations to properly describe the two main populations.  
We also find that for the same mass rocky exoplanets tend to be more irradiated than  volatile-rich exoplanets, suggesting that their H/He envelopes may have been photoevaporated by the high-energy stellar radiation. 
\\ 

The ongoing TESS mission, the future missions like CHEOPS and PLATO, and the ground-based radial velocity facilities like ESPRESSO will populate the M-R diagram with precise measurements. This will allow a better understanding  of exoplanetary demographics, in particular in the region between 2 and 4 $R_{\oplus}$, where the transition between the rocky and volatile-rich planets occurs. 
Finally, it should be noted that the M-R diagram is in fact multi-dimensional, and is affected by other parameters such as the properties of the host stars, the age of the system, etc, and is also affected by the observational biases. Therefore, when more data become available, future studies should investigate the multi-layer nature of the M-R relation, correct for the selection effects, and hopefully, provide a more complete understanding of the characteristics of planets around other stars.   

\begin{acknowledgements}
We thank the referee for valuable comments which significantly improved our paper. We also thank the Swiss National Science Foundation (SNSF), the Geneva University 
and the Zurich University for their continuous support to our exoplanet researches. We also thank George Fest for technical support with constructing the catalogue. 
This work has been in particular carried out in the frame of the National Centre 
for Competence in Research ‘PlanetS’ supported by SNSF. This research has made use of the NASA Exoplanet Archive, which is operated by the California Institute of Technology, under contract with the National Aeronautics and Space Administration under the Exoplanet Exploration Program.
\end{acknowledgements}

\bibliography{bibliography}

\appendix

\section{ }

Figure A.1 shows the evolution of the M-R diagram before and after  applying our selection criteria as discussed in Section 2.1. Figure A.2 shows density against mass for our catalogue, with the rocky population and the volatile-rich populations being divided by the  pure-water line. The grey envelope indicates the region between $2.8~g \, cm^{-3}$ to $3.3~g \, cm^{-3}$, showing that the division of the two populations by a density cutoff of $\sim 3~g \, cm^{-3}$ lead to  very similar results.
Table A.1 lists all the exoplanets up to a mass of $120 M_{\oplus}$ in our new "filtered" catalogue, and also states the references provided by the NASA Exoplanet Archive with the ones used in this work.

\begin{figure*}[h]
\centering
  \begin{tabular}{@{}cc@{}}
    \includegraphics[scale=0.47]{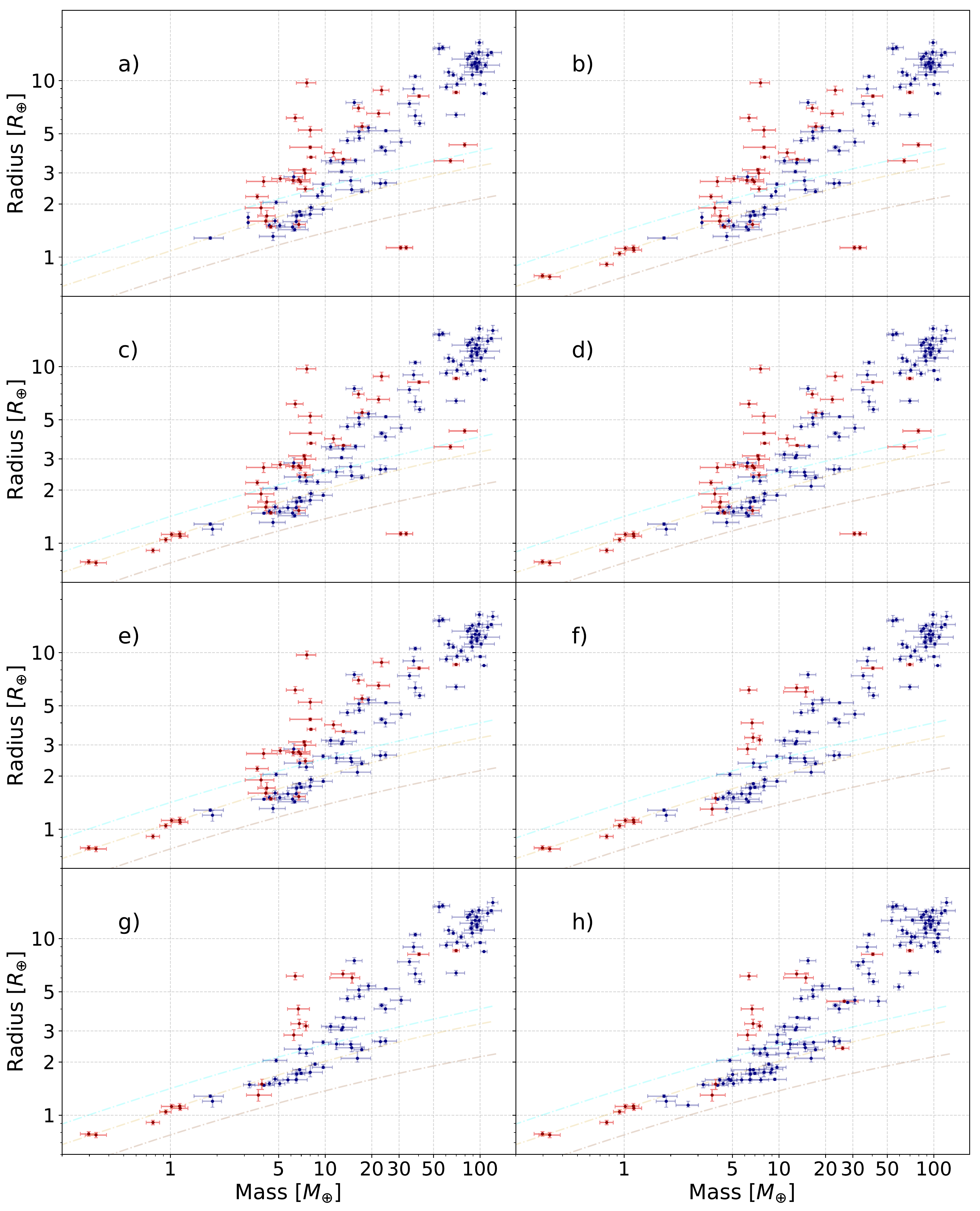}
  \end{tabular}
  \caption{The different steps of the data selection starting with initial NASA Exoplanet Archive data and ending with our final "filtered" catalogue.}
\end{figure*}

\renewcommand{\arraystretch}{1.4}

\begin{table*}[h]
\centering
\caption{Filtered catalogue with robust and reliable mass and radius measurements. It compares the reference picked by the NASA Exoplanet Archive (if available) and our selected reference. Where the radius has been obtained from \cite{Berger2018} (due to smaller uncertainty) it is indicated with an asterisk.}
\begin{tabular}{c||c|c||c|c}

\toprule

\textbf{Planet}&\textbf{Mass [$M_{\oplus}$]}&\textbf{Radius [$R_{\oplus}$]}&\textbf{Reference in NASA Exo. Arch.}&\textbf{Reference in this work}\\\hline

TRAPPIST-1 d  & $ 0.297  \ ^{ +  0.039 }  _{- 0.035 }$& $ 0.784  \ ^{ +  0.023 }  _{ - 0.023  }$&  \cite{Gillon2017}  &  \cite{grimm2018}  \\ 
TRAPPIST-1 h  & $ 0.331  \ ^{ +  0.056 }  _{- 0.049 }$& $ 0.773  \ ^{ +  0.026 }  _{ - 0.027  }$&   -   &  \cite{grimm2018}  \\ 
TRAPPIST-1 e  & $ 0.772  \ ^{ +  0.079 }  _{- 0.075 }$& $ 0.91  \ ^{ +  0.026 }  _{ - 0.027  }$&  \cite{Gillon2017}  &  \cite{grimm2018}  \\ 
TRAPPIST-1 f  & $ 0.934  \ ^{ +  0.078 }  _{- 0.08 }$& $ 1.046  \ ^{ +  0.029 }  _{ - 0.03  }$&  \cite{Gillon2017}  &  \cite{grimm2018}  \\ 
TRAPPIST-1 b  & $ 1.017  \ ^{ +  0.154 }  _{- 0.143 }$& $ 1.121  \ ^{ +  0.032 }  _{ - 0.032  }$&  \cite{Gillon2017}  &  \cite{grimm2018}  \\ 
TRAPPIST-1 g  & $ 1.148  \ ^{ +  0.098 }  _{- 0.095 }$& $ 1.127  \ ^{ +  0.041 }  _{ - 0.041  }$&  \cite{Gillon2017}  &  \cite{grimm2018} \\ 
TRAPPIST-1 c  & $ 1.156  \ ^{ +  0.142 }  _{- 0.131 }$& $ 1.095  \ ^{ +  0.031 }  _{ - 0.031  }$&  \cite{Gillon2017}  &  \cite{grimm2018}  \\ 
LHS 1140 c  & $ 1.81  \ ^{ +  0.39 }  _{- 0.39 }$& $ 1.282  \ ^{ +  0.024 }  _{ - 0.024  }$&  \cite{Ment2018} &  \cite{Ment2018}  \\
GJ 357 b  & $ 1.84  \ ^{ +  0.31 }  _{- 0.31 }$& $ 1.217  \ ^{ +  0.084 }  _{ - 0.083  }$&  -  &  \cite{Luque2019}  \\ 
Kepler-78 b  & $ 1.87  \ ^{ +  0.27 }  _{- 0.26 }$& $ 1.2  \ ^{ +  0.08 }  _{ - 0.09  }$&  \cite{stassun2017}  &  \cite{Grunblatt2015}  \\ 
K2-229 b  & $ 2.59  \ ^{ +  0.43 }  _{- 0.43 }$& $ 1.14  \ ^{ +  0.06 }  _{ - 0.03  }$&  \cite{Livingston2018}  &  \cite{Santerne2018}  \\
Kepler-10 b  & $ 3.24  \ ^{ +  0.28 }  _{- 0.28 }$& $ 1.489  \ ^{ +  0.07 }  _{ - 0.06  }$&  \cite{Esteves2015}  & \cite{Rajpaul2017}* \\ 
Kepler-80 d  & $ 3.7  \ ^{ +  0.8 }  _{- 0.6 }$& $ 1.3  \ ^{ +  0.1 }  _{ - 0.1  }$&  \cite{MacDonald2016}  & \cite{HaddenLithwick2017}  \\ 
Kepler-36 b  & $ 3.9  \ ^{ +  0.2 }  _{- 0.2 }$& $ 1.5  \ ^{ +  0.1 }  _{ - 0.1  }$&  \cite{carter2012}  &  \cite{HaddenLithwick2017}  \\ 
Kepler-93 b  & $ 4.02  \ ^{ +  0.68 }  _{- 0.68 }$& $ 1.478  \ ^{ +  0.019 }  _{ - 0.019  }$&  \cite{stassun2017}  &  \cite{Dressing2015}  \\ 
Kepler-65 d  & $ 4.14  \ ^{ +  0.79 }  _{- 0.8 }$& $ 1.587  \ ^{ +  0.04 }  _{ - 0.035  }$&  \cite{Chaplin2013}   &  \cite{Mills2019}  \\ 
HD 219134 c  & $ 4.36  \ ^{ +  0.22 }  _{- 0.22 }$& $ 1.511  \ ^{ +  0.047 }  _{ - 0.047  }$&  \cite{Gillon2017}  &  \cite{Gillon2017}  \\
HD 219134 b  & $ 4.74  \ ^{ +  0.19 }  _{- 0.19 }$& $ 1.602  \ ^{ +  0.055 }  _{ - 0.055  }$&  \cite{Gillon2017}  &  \cite{Gillon2017}  \\ 
HD 39091 c  & $ 4.82  \ ^{ +  0.84 }  _{- 0.86 }$& $ 2.042  \ ^{ +  0.05 }  _{ - 0.05  }$&   -   &  \cite{Huang2018}  \\ 
GJ 9827 b  & $ 4.89  \ ^{ +  0.477 }  _{- 0.477 }$& $ 1.575  \ ^{ +  0.03 }  _{ - 0.02  }$&  \cite{Rodriguez2018}  &  \cite{Rice2019}  \\ 
HD 3167 b  & $ 5.02  \ ^{ +  0.38 }  _{- 0.38 }$& $ 1.7  \ ^{ +  0.08 }  _{ - 0.08  }$&   -   &  \cite{Christiansen2017} \\ 
K2-141 b  & $ 5.08  \ ^{ +  0.41 }  _{- 0.41 }$& $ 1.51  \ ^{ +  0.05 }  _{ - 0.05  }$&  \cite{Malavolta2018}  & \cite{Malavolta2018} \\
CoRoT-7 b  & $ 5.74  \ ^{ +  0.86 }  _{- 0.86 }$& $ 1.585  \ ^{ +  0.064 }  _{ - 0.064  }$&  \cite{stassun2017}  &  \cite{Barros2014}  \\ 
GJ 1214 b  & $ 6.26125  \ ^{ +  0.85814 }  _{- 0.85814 }$& $ 2.847  \ ^{ +  0.202 }  _{ - 0.202  }$&  \cite{Harpsoe2013}  &  \cite{Harpsoe2013}  \\ 
K2-291 b  & $ 6.49  \ ^{ +  1.16 }  _{- 1.16 }$& $ 1.589  \ ^{ +  0.095 }  _{ - 0.072  }$&  \cite{Kosiarek2019} &  \cite{Kosiarek2019}  \\ 
K2-131 b  & $ 6.5  \ ^{ +  1.6 }  _{- 1.6 }$& $ 1.81  \ ^{ +  0.16 }  _{ - 0.12  }$&   -   &  \cite{Dai2017}  \\ 
K2-265 b  & $ 6.54  \ ^{ +  0.84 }  _{- 0.84 }$& $ 1.71  \ ^{ +  0.11 }  _{ - 0.11  }$&  \cite{Lam2018}  &  \cite{Lam2018}  \\ 
Kepler-11 e  & $ 6.7  \ ^{ +  1.2 }  _{- 1.0 }$& $ 4.0  \ ^{ +  0.2 }  _{ - 0.3  }$&  \cite{HaddenLithwick2017}  &  \cite{HaddenLithwick2017}  \\ 
Kepler-11 d  & $ 6.8  \ ^{ +  0.7 }  _{- 0.8 }$& $ 3.3  \ ^{ +  0.2 }  _{ - 0.2  }$&  \cite{Lissauer2013}  &  \cite{HaddenLithwick2017}  \\ 
WASP-47 e  & $ 6.83  \ ^{ +  0.66 }  _{- 0.66 }$& $ 1.81  \ ^{ +  0.027 }  _{ - 0.027  }$&  \cite{vanderburg2017}  &  \cite{vanderburg2017}  \\ 
Kepler-454 b  & $ 6.84  \ ^{ +  1.4 }  _{- 1.4 }$& $ 2.37  \ ^{ +  0.13 }  _{ - 0.13  }$&  \cite{stassun2017}   &  \cite{Gettel2016}  \\ 
LHS 1140 b  & $ 6.98  \ ^{ +  0.89 }  _{- 0.89 }$& $ 1.727  \ ^{ +  0.032 }  _{ - 0.032  }$&  \cite{Ment2018}  &  \cite{Ment2018}  \\ 
Kepler-36 c  & $ 7.5  \ ^{ +  0.3 }  _{- 0.3 }$& $ 3.2  \ ^{ +  0.2 }  _{ - 0.2  }$&  \cite{carter2012}  &  \cite{HaddenLithwick2017}  \\ 
HD 97658 b  & $ 7.55  \ ^{ +  0.83 }  _{- 0.79 }$& $ 2.247  \ ^{ +  0.098 }  _{ - 0.095  }$&  \cite{stassun2017} &  \cite{vangrootel2014}  \\ 
HD 15337 b  & $ 7.20  \ ^{ +  0.81 }  _{- 0.81 }$& $ 1.70  \ ^{ +  0.06 }  _{ - 0.06  }$&  \cite{Gandolfi2019}  &  \cite{Dumusque2019}  \\ 
K2-216 b  & $ 8.0  \ ^{ +  1.6 }  _{- 1.6 }$& $ 1.75  \ ^{ +  0.17 }  _{ - 0.1  }$&  \cite{persson2018}  &  \cite{persson2018}  \\ 
Kepler-19 b  & $ 8.4  \ ^{ +  1.6 }  _{- 1.5 }$& $ 2.2  \ ^{ +  0.07 }  _{ - 0.07  }$&  \cite{Malavolta2018}   &  \cite{Malavolta2018}  \\ 
55 Cnc e  & $ 8.59  \ ^{ +  0.43 }  _{- 0.43 }$& $ 1.947  \ ^{ +  0.038 }  _{ - 0.038  }$&  \cite{Demory2016}  &  \cite{Crida2018}  \\ 
...&...&...&...&...\\

\end{tabular}

\end{table*}

\begin{table*}[h]
\centering

\begin{tabular}{c||cc||c|c}
\toprule

\textbf{Planet}&\textbf{Mass [$M_{\oplus}$]}&\textbf{Radius [$R_{\oplus}$]}&\textbf{Reference in NASA Exo. Arch.}&\textbf{Reference in this work}\\\hline

HD 15337 c  & $ 8.79  \ ^{ +  1.68 }  _{- 1.68 }$& $ 2.52  \ ^{ +  0.11 }  _{ - 0.11  }$&   -   &  \cite{Dumusque2019}  \\ 
HD 213885 b  & $ 8.83  \ ^{ +  0.66 }  _{- 0.65 }$& $ 1.745  \ ^{ +  0.051 }  _{ - 0.052  }$&   -   &  \cite{Espinoza2019}  \\ 
EPIC 220674823 b  & $ 9.0  \ ^{ +  1.6 }  _{- 1.6 }$& $ 1.82  \ ^{ +  0.1 }  _{ - 0.1  }$&   -   &  \cite{Sinukoff2017}  \\ 
Kepler-107 c  & $ 9.39  \ ^{ +  1.77 }  _{- 1.77 }$& $ 1.597  \ ^{ +  0.026 }  _{ - 0.026  }$&   \cite{Bonomo2019}   &  \cite{Bonomo2019}  \\ 
K2-285 b  & $ 9.68  \ ^{ +  1.2 }  _{- 1.3 }$& $ 2.59  \ ^{ +  0.06 }  _{ - 0.06  }$&  \cite{Palle2019}  &  \cite{Palle2019}  \\ 
Kepler-20 b  & $ 9.7  \ ^{ +  1.41 }  _{- 1.44 }$& $ 1.868  \ ^{ +  0.066 }  _{ - 0.034  }$&  \cite{Buchhave2016}  &  \cite{Buchhave2016} \\ 
HD 3167 c  & $ 9.8  \ ^{ +  1.3 }  _{- 1.23 }$& $ 2.86  \ ^{ +  0.22 }  _{ - 0.22  }$&   -   &  \cite{Christiansen2017}  \\ 
Kepler-94 b  & $ 10.84  \ ^{ +  1.4 }  _{- 1.4 }$& $ 3.186  \ ^{ +  0.13 }  _{ - 0.25  }$&  \cite{Marcy2014}  &  \cite{Marcy2014}*  \\ 
K2-180 b  & $ 11.448  \ ^{ +  1.9 }  _{- 1.9 }$& $ 2.24  \ ^{ +  0.12 }  _{ - 0.12  }$&   -   &  \cite{Korth2019}  \\ 
HIP 116454 b  & $ 11.82  \ ^{ +  1.33 }  _{- 1.33 }$& $ 2.53  \ ^{ +  0.18 }  _{ - 0.18  }$&  \cite{stassun2017}  &  \cite{vanderburg2017}  \\ 
Kepler-20 c  & $ 12.75  \ ^{ +  2.17 }  _{- 2.24 }$& $ 3.047  \ ^{ +  0.064 }  _{ - 0.056  }$&  \cite{Buchhave2016}  &  \cite{Buchhave2016}  \\ 
Kepler-95 b  & $ 13.0  \ ^{ +  2.9 }  _{- 2.9 }$& $ 3.145  \ ^{ +  0.144 }  _{ - 0.132  }$&  \cite{Marcy2014}  &  \cite{Marcy2014}*  \\ 
KOI-94 e  & $ 13.0  \ ^{ +  2.5 }  _{- 2.1 }$& $ 6.31  \ ^{ +  0.3 }  _{ - 0.3  }$&  \cite{weiss2013}  &  \cite{Masuda2013}*  \\ 
WASP-47 d  & $ 13.1  \ ^{ +  1.5 }  _{- 1.5 }$& $ 3.576  \ ^{ +  0.046 }  _{ - 0.046  }$&  \cite{vanderburg2017}  &  \cite{vanderburg2017}  \\ 
GJ 3470 b  & $ 13.9  \ ^{ +  1.5 }  _{- 1.5 }$& $ 4.57  \ ^{ +  0.18 }  _{ - 0.18  }$&  \cite{Awiphan2016}  &  \cite{Awiphan2016}  \\ 
Kepler-48 c  & $ 14.61  \ ^{ +  2.3 }  _{- 2.3 }$& $ 2.522  \ ^{ +  0.113 }  _{ - 0.107  }$&  \cite{Marcy2014}  &  \cite{Marcy2014}*  \\ 
K2-263 b  & $ 14.8  \ ^{ +  3.1 }  _{- 3.1 }$& $ 2.41  \ ^{ +  0.12 }  _{ - 0.12  }$&  \cite{Mortier2018}  &  \cite{Mortier2018}  \\ 
Kepler-18 d  & $ 14.9  \ ^{ +  1.8 }  _{- 4.2 }$& $ 6.0  \ ^{ +  0.4 }  _{ - 0.4  }$&  \cite{Cochran2011}  &  \cite{HaddenLithwick2017}  \\ 
K2-24 c  & $ 15.4  \ ^{ +  1.9 }  _{- 1.8 }$& $ 7.5  \ ^{ +  0.3 }  _{ - 0.2  }$&   -   &  \cite{Petigura2018}  \\ 
K2-285 c  & $ 15.68  \ ^{ +  2.28 }  _{- 2.13 }$& $ 3.53  \ ^{ +  0.08 }  _{ - 0.08  }$&  \cite{Petigura2018}  &  \cite{Petigura2018}  \\ 
Kepler-131 b  & $ 16.13  \ ^{ +  3.5 }  _{- 3.5 }$& $ 2.1  \ ^{ +  0.2 }  _{ - 0.1  }$&  \cite{Marcy2014}  &  \cite{Marcy2014}*  \\ 
K2-32 b  & $ 16.5  \ ^{ +  2.7 }  _{- 2.7 }$& $ 5.13  \ ^{ +  0.28 }  _{ - 0.28  }$&  \cite{Petigura2017}  & \cite{Petigura2017}  \\ 
HD 219666 b  & $ 16.6  \ ^{ +  1.3 }  _{- 1.3 }$& $ 4.71  \ ^{ +  0.17 }  _{ - 0.17  }$&  \cite{Esposito2018}  &  \cite{Esposito2018}  \\ 
K2-110 b  & $ 16.7  \ ^{ +  3.2 }  _{- 3.2 }$& $ 2.59  \ ^{ +  0.01 }  _{ - 0.01  }$&  \cite{Osborn2017} &  \cite{Osborn2017}  \\ 
Kepler-10 c  & $ 17.2  \ ^{ +  1.9 }  _{- 1.9 }$& $ 2.35  \ ^{ +  0.09 }  _{ - 0.04  }$&  \cite{Dumusque2014}  &  \cite{Dumusque2014}  \\ 
K2-24 b  & $ 19.0  \ ^{ +  -2.1 }  _{- -2.0 }$& $ 5.4  \ ^{ +  0.2 }  _{ - 0.2  }$&  \cite{Petigura2018}  &  \cite{Petigura2018}  \\ 
GJ 143 b  & $ 22.7  \ ^{ +  2.2 }  _{- 1.9 }$& $ 2.61  \ ^{ +  0.17 }  _{ - 0.16  }$&   -   &  \cite{Dragomir2019}  \\ 
GJ 436 b  & $ 23.1  \ ^{ +  0.8 }  _{- 0.8 }$& $ 4.191  \ ^{ +  0.1 }  _{ - 0.1  }$&  \cite{Turner2016}  &  \cite{Turner2016}  \\ 
HD 21749 b  & $ 23.2  \ ^{ +  2.13 }  _{- 1.91 }$& $ 2.84  \ ^{ +  0.16 }  _{ - 0.16  }$&   -   &  \cite{Dragomir2019}  \\ 
Kepler-4 b  & $ 24.472  \ ^{ +  3.814 }  _{- 3.814 }$& $ 4.002  \ ^{ +  0.213 }  _{ - 0.213  }$& \cite{Borucki2010}  &  \cite{Borucki2010}  \\ 
HD 119130 b  & $ 24.5  \ ^{ +  4.4 }  _{- 4.4 }$& $ 2.63  \ ^{ +  0.11 }  _{ - 0.1  }$&  \cite{Luque2018}  &  \cite{Luque2018}  \\ 
Kepler-25 c  & $ 24.6  \ ^{ +  5.7 }  _{- 5.7 }$& $ 5.154  \ ^{ +  0.06 }  _{ - 0.06  }$&  \cite{Marcy2014}  &  \cite{Marcy2014}*  \\ 
Kepler-411 b  & $ 25.758  \ ^{ +  2.544 }  _{- 2.544 }$& $ 2.3968  \ ^{ +  0.056 }  _{ - 0.056  }$&  \cite{Sun2019}   &  \cite{Sun2019}  \\ 
Kepler-411 c  & $ 26.394  \ ^{ +  6.042 }  _{- 6.042 }$& $ 4.418  \ ^{ +  0.06 }  _{ - 0.06  }$&   \cite{Sun2019}   &  \cite{Sun2019}  \\ 
HAT-P-11 b  & $ 27.76  \ ^{ +  3.08 }  _{- 3.08 }$& $ 4.35  \ ^{ +  0.05 }  _{ - 0.05  }$&  \cite{Yee2018}  &  \cite{Allart2018}  \\ 
K2-27 b  & $ 30.9  \ ^{ +  4.6 }  _{- 4.6 }$& $ 4.48  \ ^{ +  0.23 }  _{ - 0.23  }$&  \cite{Petigura2017}  &  \cite{Petigura2017}  \\ 
WASP-166 b  & $ 32.436  \ ^{ +  1.272 }  _{- 1.272 }$& $ 7.056  \ ^{ +  0.336 }  _{ - 0.336  }$&   -   &  \cite{Hellier2017}  \\ 
HD 89345 b  & $ 34.9613  \ ^{ +  5.40311 }  _{- 5.72094 }$& $ 7.398  \ ^{ +  0.314 }  _{ - 0.336  }$&  \cite{Yu2018}  &  \cite{Yu2018}  \\ 
WASP-139 b  & $ 37.18611  \ ^{ +  5.40311 }  _{- 5.40311 }$& $ 8.967  \ ^{ +  0.56 }  _{ - 0.56  }$&  \cite{Hellier2017}  &  \cite{Hellier2017}  \\ 
WASP-107 b  & $ 38.0  \ ^{ +  3.0 }  _{- 3.0 }$& $ 10.6  \ ^{ +  0.3 }  _{ - 0.3  }$&   -   &  \cite{Anderson2017}  \\ 
HATS-7 b  & $ 38.1  \ ^{ +  3.8 }  _{- 3.8 }$& $ 6.31  \ ^{ +  0.5 }  _{ - 0.4  }$&  \cite{Bakos2015}  &  \cite{Bakos2015}  \\ 
...&...&...&...&...\\

\end{tabular}

\end{table*}
   
\begin{table*}[h]
\centering

\begin{tabular}{c||cc||c|c}
\toprule

\textbf{Planet}&\textbf{Mass [$M_{\oplus}$]}&\textbf{Radius [$R_{\oplus}$]}&\textbf{Reference}\\\hline

Kepler-35 b  & $ 40.363  \ ^{ +  6.356 }  _{- 6.356 }$& $ 8.16  \ ^{ +  0.157 }  _{ - 0.157  }$&  \cite{Welsh2012}  &  \cite{Welsh2012}  \\ 
WASP-156 b  & $ 40.68224  \ ^{ +  3.1783 }  _{- 2.86047 }$& $ 5.717  \ ^{ +  0.224 }  _{ - 0.224  }$&  \cite{Demangeon2017}  &  \cite{Demangeon2017}  \\ 
K2-55 b  & $ 43.88  \ ^{ +  5.4 }  _{- 5.4 }$& $ 4.424  \ ^{ +  0.29 }  _{ - 0.29  }$&   \cite{Crossfield2016}&  \cite{Dressing2018}  \\ 
Kepler-101 b  & $ 51.1  \ ^{ +  5.1 }  _{- 4.7 }$& $ 5.986  \ ^{ +  0.27 }  _{ - 0.25  }$&  \cite{Bonomo2014}  &  \cite{Bonomo2014}*  \\ 
HAT-P-48 b  & $ 53.4  \ ^{ +  7.6 }  _{- 7.6 }$& $ 12.66  \ ^{ +  0.6 }  _{ - 0.6  }$&   -   &  \cite{Bakos2016}  \\ 
KELT-11 b  & $ 54.3  \ ^{ +  4.8 }  _{- 4.8 }$& $ 15.1  \ ^{ +  1.1 }  _{ - 1.1  }$&  \cite{Beatty2017}  &  \cite{Beatty2017}  \\ 
K2-261 b  & $ 56.922  \ ^{ +  6.36 }  _{- 6.36 }$& $ 9.4  \ ^{ +  0.12 }  _{ - 0.12  }$&  \cite{Johnson2018}  &  \cite{Brahm2019}  \\ 
WASP-127 b  & $ 57.2094  \ ^{ +  6.3566 }  _{- 6.3566 }$& $ 15.356  \ ^{ +  0.448 }  _{ - 0.448  }$&  \cite{Lam2017}  &  \cite{Lam2017}  \\ 
K2-108 b  & $ 59.4  \ ^{ +  4.4 }  _{- 4.4 }$& $ 5.33  \ ^{ +  0.21 }  _{ - 0.21  }$&   \cite{Petigura2017} &  \cite{Petigura2017}  \\ 
HAT-P-18 b  & $ 62.3  \ ^{ +  2.5 }  _{- 2.5 }$& $ 11.153  \ ^{ +  0.583 }  _{ - 0.583  }$&  \cite{Hartman2011}  &  \cite{Esposito2014}  \\ 
HD 221416 b  & $ 63.4  \ ^{ +  5.7 }  _{- 5.7 }$& $ 9.16  \ ^{ +  0.34 }  _{ - 0.31  }$&  \cite{Huber2019}  &  \cite{Huber2019}  \\ 
HAT-P-47 b  & $ 65.508  \ ^{ +  12.4 }  _{- 12.4 }$& $ 14.7  \ ^{ +  0.4 }  _{ - 0.4  }$&   -   &  \cite{Bakos2016}  \\ 
HAT-P-12 b  & $ 67.059  \ ^{ +  3.814 }  _{- 3.814 }$& $ 10.749  \ ^{ +  0.325 }  _{ - 0.235  }$&   \cite{Hartman2009} & \cite{Huber2019}   \\ 
CoRoT-8 b  & $ 69.92  \ ^{ +  9.53 }  _{- 9.53 }$& $ 6.39  \ ^{ +  0.22 }  _{ - 0.22  }$&  \cite{Borde2010}  &  \cite{Borde2010}  \\ 
Kepler-34 b  & $ 69.92  \ ^{ +  3.496 }  _{- 3.178 }$& $ 8.564  \ ^{ +  0.135 }  _{ - 0.157  }$&  \cite{Welsh2012}  &  \cite{Welsh2012}  \\ 
Kepler-425 b  & $ 71.8  \ ^{ +  14.6 }  _{- 14.6 }$& $ 10.255  \ ^{ +  0.47 }  _{ - 0.45  }$&   -   &  \cite{bonomo2017}* \\ 
NGTS-5 b  & $ 72.8  \ ^{ +  11.8 }  _{- 11.8 }$& $ 12.73  \ ^{ +  0.26 }  _{ - 0.26  }$&   -   &  \cite{Eigmuller2014}  \\ 
HATS-5 b  & $ 75.323  \ ^{ +  3.814 }  _{- 3.914 }$& $ 10.223  \ ^{ +  0.28 }  _{ - 0.28  }$&  \cite{Zhou2014}  &  \cite{Zhou2014}  \\ 
WASP-29 b  & $ 77.9  \ ^{ +  7.3 }  _{- 7.0 }$& $ 8.8  \ ^{ +  0.6 }  _{ - 0.4  }$&  \cite{stassun2017}  &  \cite{gibson2013}  \\ 
WASP-69 b  & $ 82.632  \ ^{ +  5.4 }  _{- 5.4 }$& $ 9.11  \ ^{ +  0.3 }  _{ - 0.3  }$&  \cite{stassun2017}  &  \cite{anderson2013}  \\ 
HATS-43 b  & $ 83.0  \ ^{ +  17.0 }  _{- 17.0 }$& $ 13.23  \ ^{ +  0.56 }  _{ - 0.56  }$&   -   &  \cite{Brahm2019}  \\ 
WASP-131 b  & $ 85.8141  \ ^{ +  6.3566 }  _{- 6.3566 }$& $ 13.675  \ ^{ +  0.56 }  _{ - 0.56  }$&  \cite{Hellier2017}  &  \cite{Hellier2017} \\ 
WASP-117 b  & $ 87.55  \ ^{ +  2.86 }  _{- 2.8 }$& $ 11.44  \ ^{ +  0.785 }  _{ - 0.785  }$&  \cite{stassun2017}  &  \cite{lendl2014}  \\ 
WASP-160 b  & $ 88.35674  \ ^{ +  13.98452 }  _{- 14.30235 }$& $ 12.218  \ ^{ +  0.527 }  _{ - 0.46  }$&   -   &  \cite{Lendl2019}  \\ 
WASP-39 b  & $ 88.989  \ ^{ +  9.535 }  _{- 9.535 }$& $ 14.235  \ ^{ +  0.448 }  _{ - 0.448  }$&  \cite{Faedi2011}  &  \cite{Faedi2011}  \\ 
WASP-126 b  & $ 88.9924  \ ^{ +  12.7132 }  _{- 12.7132 }$& $ 10.761  \ ^{ +  1.121 }  _{ - 0.56  }$&  \cite{Maxted2016}  &  \cite{Maxted2016}  \\ 
Kepler-427 b  & $ 92.8  \ ^{ +  17.2 }  _{- 17.2 }$& $ 12.696  \ ^{ +  0.577 }  _{ - 0.547  }$&   -   &  \cite{bonomo2017}*  \\ 
HAT-P-19 b  & $ 92.802  \ ^{ +  5.721 }  _{- 5.721 }$& $ 12.689  \ ^{ +  0.807 }  _{ - 0.807  }$&  \cite{Hartman2011}  &  \cite{Hartman2011}  \\ 
WASP-181 b  & $ 95.0  \ ^{ +  10.8 }  _{- 10.8 }$& $ 13.26  \ ^{ +  0.66 }  _{ - 0.79  }$&   -   &  \cite{Turner2019}  \\ 
WASP-21 b  & $ 95.345  \ ^{ +  3.496 }  _{- 3.496 }$& $ 11.99  \ ^{ +  0.56 }  _{ - 0.56  }$&  \cite{Bouchy2010}  &  \cite{Bouchy2010}  \\ 
WASP-83 b  & $ 95.349  \ ^{ +  9.5349 }  _{- 9.5349 }$& $ 11.657  \ ^{ +  0.897 }  _{ - 0.56  }$&  \cite{Hellier2015}  &  \cite{Hellier2015}  \\ 
HAT-P-51 b  & $ 98.20947  \ ^{ +  5.72094 }  _{- 5.72094 }$& $ 14.493  \ ^{ +  0.605 }  _{ - 0.605  }$&  \cite{Hartman2015}  &  \cite{Hartman2015}  \\ 
WASP-151 b  & $ 98.5273  \ ^{ +  12.7132 }  _{- 9.5349 }$& $ 12.666  \ ^{ +  0.336 }  _{ - 0.336  }$&  \cite{Demangeon2017}  &  \cite{Demangeon2017}  \\ 
WASP-20 b  & $ 98.8  \ ^{ +  6.0 }  _{- 5.8 }$& $ 16.39  \ ^{ +  0.66 }  _{ - 0.66  }$&  \cite{Anderson2015}  &  \cite{bonomo2017}  \\ 
K2-287 b  & $ 100.0  \ ^{ +  9.0 }  _{- 9.0 }$& $ 9.49  \ ^{ +  0.15 }  _{ - 0.15  }$&  \cite{Jordan2019}  &  \cite{Jordan2019}  \\ 
HATS-6 b  & $ 101.0  \ ^{ +  22.0 }  _{- 22.0 }$& $ 11.19  \ ^{ +  0.21 }  _{ - 0.21  }$&  \cite{Hartman2015}  &  \cite{Hartman2015}  \\ 
HD 149026 b  & $ 102.0  \ ^{ +  4.0 }  _{- 4.0 }$& $ 9.09  \ ^{ +  0.3 }  _{ - 0.3  }$&  \cite{stassun2017}  &  \cite{bonomo2017}  \\ 
Kepler-16 b  & $ 105.833  \ ^{ +  5.085 }  _{- 5.085 }$& $ 8.449  \ ^{ +  0.029 }  _{ - 0.029  }$&  \cite{Doyle2011}  &  \cite{Doyle2011}  \\ 
K2-295 b & $ 106.0  \ ^{ +  20.0 }  _{- 20.0 }$& $ 10.1  \ ^{ +  0.1 }  _{ - 0.1  }$&   -   &  \cite{Smith2018}  \\ 
EPIC 220501947 b  & $ 106.848  \ ^{ +  3.816 }  _{- 3.816 }$& $ 10.6064  \ ^{ +  0.056 }  _{ - 0.1344  }$&   -   &  \cite{Smith2018-2}  \\ 
Kepler-426 b  & $ 107.0  \ ^{ +  18.0 }  _{- 19.0 }$& $ 12.22  \ ^{ +  0.34 }  _{ - 0.34  }$&   -   &  \cite{bonomo2017}  \\ 
HAT-P-44 b  & $ 111.871  \ ^{ +  9.217 }  _{- 9.217 }$& $ 13.922  \ ^{ +  1.188 }  _{ - 0.572  }$&  \cite{Hartman2014}  &  \cite{Hartman2014} \\ 
Qatar-8 b  & $ 117.978  \ ^{ +  19.716 }  _{- 19.716 }$& $ 14.392  \ ^{ +  0.2464 }  _{ - 0.2464  }$&   -   &  \cite{Alsubai2019} \\ 
WASP-63 b  & $ 120.77  \ ^{ +  9.5 }  _{- 9.5 }$& $ 16.0  \ ^{ +  1.1 }  _{ - 0.7  }$&  \cite{stassun2017}  &  \cite{hellier2012}  \\

\hline
\end{tabular}

\end{table*}  

 \begin{figure*}[h]
\centering
  \begin{tabular}{@{}cc@{}}
    \includegraphics[scale=0.45]{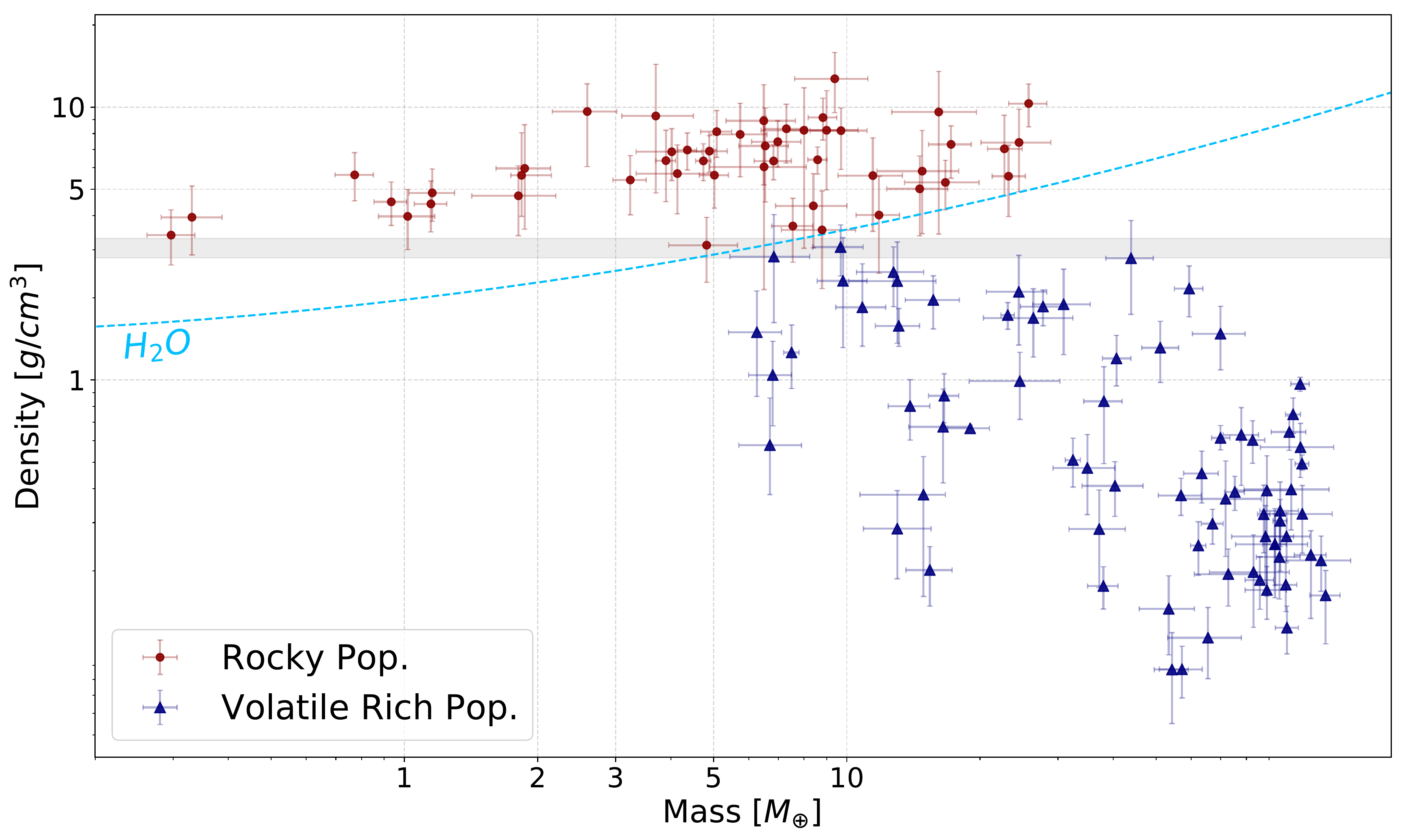}
  \end{tabular}
  \caption{Density vs.~mass for our revised catalogue. The rocky population and the volatile-rich population are separated by the composition line of pure water \cite[][]{Dorn2015}. The grey envelope indicates the region between $2.8~g \, cm^{-3}$ and $3.3~g \, cm^{-3}$. }.
\end{figure*}

\end{document}